\documentclass[12pt]{article}
\usepackage{graphicx}
\usepackage{amsmath}
\usepackage{amsfonts}
\usepackage{amssymb} 
\usepackage{graphicx} 
\usepackage{cite} 
\newcommand{\im}{{\rm Im}}

\newcommand{\be}{\begin{equation}}
\newcommand{\ee}{\end{equation}}
\newcommand{\bea}{\begin{eqnarray}}
\newcommand{\eea}{\end{eqnarray}}
\newcommand{\beq}{\begin{equation}}
\newcommand{\eeq}{\end{equation}}
\newcommand{\ba}{\begin{array}}
\newcommand{\ea}{\end{array}}
\newcommand{\beqa}{\begin{eqnarray}}
\newcommand{\eeqa}{\end{eqnarray}}

\newcommand{\cL}{{\cal L}}

\newcommand{\no}{\nonumber}

\newcommand{\lsim}{\stackrel{<}{_\sim}}
\newcommand{\gsim}{\stackrel{>}{_\sim}}

\newcommand\dll{\delta_{LL}^{d}} \newcommand\drr{\delta_{RR}^{d}}

\newcommand\tanb{\tan\beta}

\newcommand\epsg{\epsilon_s}

\newcommand{\newc}{\newcommand}
\newc\eg{{\it {e.g.}}}  \newc\vs{{\it {vs.}}}   \newc\etal{{\it {et al.}}}
\newc\etc{{\it {etc.}}} \newc\ie{{\it {i.e.}}}

\def \text{\mathrm}

\begin{document}

\begin{flushright}
\hfill \today \\
\end{flushright}
\vskip   1 true cm 
\begin{center}
{\Large \textbf{New Two-loop Contributions to Hadronic EDMs\\ in the MSSM}}
\\ [20 pt]
\textsc{Junji Hisano}${}^{1}$, \textsc{Minoru Nagai}${}^{1}$
and \textsc{Paride Paradisi}${}^{2,3}$
  \\ [20 pt]
${}^{1}~$\textsl{ICRR, University of Tokyo, Kashiwa 277-8582, Japan } \\ 
[5pt]
${}^{2}~$\textsl{ INFN, Sezione di Roma II and Dipartimento di Fisica, 
\\ Universit\`a di Roma
``Tor Vergata'', I-00133 Rome, Italy}\\ 
[5pt]
${}^{3}~$\textsl{Department of Physics, Technion-Israel Institute of Technology,\\
Technion City, 32000 Haifa, Israel}

\vskip   1 true cm

\textbf{Abstract\\}
\end{center}
\noindent
Flavor-changing terms with CP-violating phases in the quark sector may
contribute to the hadronic electric dipole moments (EDMs). However,
within the Standard Model (SM), the source of CP violation comes from
the unique CKM phase, and it turns out that the EDMs are strongly
suppressed.  This implies that the EDMs are very sensitive to
non-minimal flavor violation structures of theories beyond the SM.  In
this paper, we discuss the quark EDMs and CEDMs (chromoelectric dipole
moments) in the MSSM with general flavor-changing terms in the squark
mass matrices. In particular, the charged-Higgs mediated contributions
to the down-quark EDM and CEDM are evaluated at two-loop level.  We
point out that these two-loop contributions may dominate over the
one-loop induced gluino or Higgsino contributions even when the squark
and gluino masses are around few TeV and $\tan\beta$ is moderate.

\vskip   1 true cm

\section{Introduction}

Within the Standard Model (SM), the sources of CP violation are the
phase in the CKM matrix and the QCD theta $(\overline{\theta})$
parameter. The former induces CP violation in flavor-changing
processes, such as CP-violating $K$ and $B$ meson decay modes, whereas
the latter induces flavor-conserving CP violation, such as neutron
electric dipole moment (EDM).  The experimental upper bound on  the
neutron EDM gives a strong constraint on $\overline{\theta}$, as
$|\overline{\theta}| \lsim 10^{-(10-11)}$.  On the other hand, the
recent measurements of CP asymmetries in the $B$ decay modes at the
Babar and Belle experiments confirm that the CKM phase is almost
maximal and it is the dominant source of CP violation in the $K$ and
$B$ meson decays.

The minimal supersymmetric SM (MSSM) is the most motivated model
beyond the SM. This model exhibits plenty of new CP-violating phases
in addition to the unique CKM phase. It is known that they constitute
a threat for very sensitive CP tests, at least for SUSY particle
masses which are within the TeV region.  New CP phases may be
introduced in both the flavor-conserving and the flavor-changing soft
SUSY breaking terms. The former ones are constrained directly by
the electron and hadronic EDMs. In the latter case, flavor-changing
neutral current (FCNC) processes are sensitive to those new
phases, and we have  to satisfy the stringent bounds arising from
the $K$ and $B$ physics low-energy experiments.

The strong suppression of the EDMs within the SM and their high
sensitivities to physics beyond the SM signal the unique possibilities
offered by the EDMs in probing the underlying mechanism of CP
violation.  In the present paper, we are interested in the quark EDMs
and CEDMs (chromoelectric dipole moments) arising from flavor-changing
terms in the squark mass matrices. While the EDMs and CEDMs are highly
suppressed within the minimal flavor violation (MFV) flamework, as it
happens in the SM, those flavor-changing terms allow us to introduce
new types of Jarlskog invariants in the MSSM.  The Jarlskog invariants
are a basis-independent measure of CP violation
\cite{Jarlskog:1985ht}. Within a general flavor violation (GFV)
framework, the invariants are less suppressed by the Yukawa coupling
constants compared to the MFV case. This implies that the EDMs are
sensitive to non-minimal structures of flavor violation.  In this
paper, we will analyze in detail two-loop effects induced by
non-holomorphic Yukawa interactions \cite{HRS,Babu:1999hn} through the
effective couplings of down-type quarks/charged-Higgs boson/up-type quarks.
Finally, we will also briefly discuss the correlations between quarks
CEDMs and $B$ physics observables.

\section{Hadronic EDMs and New Jarlskog Invariants in the MSSM}

The effective CP-violating interactions, which contribute to
the hadronic EDMs, up to dimension five operators, are
\begin{eqnarray}
{\cal L}_{CP}=\overline{\theta} \frac{\alpha_s}{8\pi} G\tilde{G}
-\sum_{q=u,d,s} i \frac{d_q}{2} 
\bar{q} (F \cdot \sigma) \gamma_5 q
-
\sum_{q=u,d,s} i\frac{d^c_q}2 \bar{q}(g_s G\cdot \sigma) \gamma_5 q,
\label{eff_cp}
\end{eqnarray}
where $F_{\mu\nu}$ and $G_{\mu\nu}$ are the electromagnetic and the
SU(3)$_C$ gauge field strengths, respectively \cite{Pospelov:2005pr}.  The
first term in Eq.~(\ref{eff_cp}) is the effective QCD theta term,
while $d_q$ and $d_q^c$ are the quark EDM and CEDM, respectively.
Within the SM, the sources of CP violation come from the QCD theta
term and the CKM matrix, $V$.  The theta term is strongly constrained
by the experimental bounds on the neutron EDM, as mentioned in the
Introduction.  One of the most natural ways to achieve such a
suppression is to impose a Peccei-Quinn symmetry \cite{Peccei:1977hh},
since the axion field makes $\bar{\theta}$ dynamically
vanishing.

Under the above assumption, the hadronic EDMs are strongly suppressed
in the SM.  The EDMs and CEDMs of the $i$-th down- and up-type quarks are
proportional to the flavor-conserving Jarlskog invariants,
\begin{eqnarray}
J^{(d_i)}_{\rm SM}= {\rm Im} \left\{Y_d[Y_d, Y_u]Y_u f_d\right\}_{ii},
\nonumber\\    
J^{(u_i)}_{\rm SM}= {\rm Im} \left\{Y_d[Y_d, Y_u]Y_u f_u\right\}_{ii},
\end{eqnarray}
respectively. Here, $(f_d)_{ij}$ [$(f_u)_{ij}$] is the down[up]-type
quark Yukawa coupling constant and $Y_{d(u)} \equiv
f_{d(u)}f_{d(u)}^\dagger$.  The suffixes $i(j)$ in the Yukawa coupling
constants refer to left(right)-handed quarks. Since the Jarlskog
invariants are of the ninth order in the Yukawa coupling constants,
the quark (C)EDMs are highly suppressed.  The first non-vanishing
contributions come from terms of order $O(\alpha_sG_F^2)$, and $d_q/e$
and $d_q^c$ are $\sim 10^{-(33-34)}~{\rm cm}$.  The neutron EDM may be
enhanced by long-distance effects, however, the SM predictions for the
hadronic EDMs, including the neutron EDM, are well below the actual
and expected future experimental resolution.

The strong suppression of EDMs within a SM framework with the
Peccei-Quinn symmetry active, comes from the fact that the CKM
contains only one physical phase that is associated, in addition, to
the only flavor-changing couplings.  However, the above situation may
be no longer valid in theories beyond the SM.  For instance, within a
SUSY framework, various CP-violating couplings may appear after the
introduction of SUSY breaking terms.  The first ones are the $F$-term
SUSY breaking terms, such as the gaugino masses, the $B$ term in the
Higgs potential and the $A$ terms for trilinear scalar couplings.  The
relative phases of the above SUSY breaking terms contribute to the
hadronic and electronic EDMs, and the experimental bounds on the EDMs
infer strong constraints on their sizes \cite{susycp}.  For instance,
barring accidental cancellations among contributions from different
phases, it turns out that
\beq
|\sin\phi_A|\lsim \left(\frac{m_{SUSY}}{\rm{TeV}}\right)^2\,,
\qquad
|\sin\phi_{B}|\lsim 
\left(\frac{m_{SUSY}}{\rm{TeV}}\right)^2\frac{1}{\tan\beta}.
\eeq
Here, $\phi_{A/B}$ is the relative phase  between the $A/B$ parameters and
the gaugino masses, and $m_{SUSY}$ is the typical SUSY particle mass
scale.

The problem of so small CP-violating phases is referred to as the SUSY
CP problem.  Here, we simply assume that the $F$-term SUSY breaking terms
are real.  Even under this assumption, we might still have CP-violating
phases in the $D$-term SUSY breaking terms, which are sfermion mass
terms.  The off-diagonal (flavor-changing) terms in the sfermion mass
matrices may have CP phases.  In the SM, the Jarlskog invariants
contributing to the hadronic EDMs are highly suppressed by the
structure of the MFV, in which flavor and CP violations come from 
the only interplays between up- and down-type quark Yukawa couplings.
However, in the MSSM we can introduce new invariants even
under the assumption that the $F$-term SUSY breaking terms are real.
Non-minimal flavor structures in the $D$-terms contribute to new
invariants that are, finally, no more suppressed by high powers of
the Yukawa coupling constants as it happens, on the contrary, in the SM.
This implies that, the hadronic EDMs measurements probe non-minimal
flavor structures in the MSSM.

The size and pattern of these flavor violation sources in the sfermion
mass matrices depend on the SUSY breaking mechanism and interactions
of the high-energy theories beyond the MSSM. In the following, 
the GFV is assumed in the squark mass terms, and we introduce, as usual, 
the mass insertion parameters defined as
\begin{eqnarray}
(\delta_{LL}^q)_{ij}\equiv 
\frac{(m^2_{\tilde{q}_L})_{ij}}{\overline{m}_{\tilde{q}_L}^2},\nonumber\\
 (\delta_{RR}^d)_{ij}\equiv
\frac{(m^2_{\tilde{d}_R})_{ij}}{\overline{m}_{\tilde{d}_R}^2},\nonumber\\
 (\delta_{RR}^u)_{ij}\equiv 
\frac{(m^2_{\tilde{u}_R})_{ij}}{\overline{m}_{\tilde{u}_R}^2},
\end{eqnarray}
for $i\ne j$. The diagonal components of the mass insertion parameters
are zero. When calculating the SUSY contribution to the quark EDMs and CEDMs
explicitly, we will take a basis in which the down-type quark Yukawa
coupling constants are flavor-diagonal. In this case,
$\delta_{LL}^d=\delta_{LL}^q$. In the flavor basis, the mass insertion
parameters for left-handed up- and down-type squarks are related by the
SU(2)$_{L}$ symmetry as $\delta_{LL}^u=V\delta_{LL}^dV^\dagger$.

In the MSUGRA model, the renormalization-group (RG) effect induced by
the top-quark Yukawa coupling generates sizable flavor-changing terms
in the left-handed down-type squark mass matrix due to the CKM mixing,
and it turns out that ${(\delta^{d}_{LL})_{ij}}\propto
-{V_{3i}^{*}}{V_{3j}}$ for moderate $\tan\beta$ values.  In the case,
the $\delta_{LL}^u$ mass insertions are close to zero.

First, let us consider a case of $(\delta_{LL}^q)_{ij}\ne 0$. The
new flavor-conserving Jarlskog invariants, which contribute to the
quark EDMs and CEDMs, appear at the third order in the Yukawa coupling
constants,
\begin{eqnarray}
J^{(d_i)}_{LL}= {\rm Im} \left\{[Y_u,\delta_{LL}^q]f_d\right\}_{ii},
\nonumber\\
J^{(u_i)}_{LL}= {\rm Im} \left\{[Y_d,\delta_{LL}^q]f_u\right\}_{ii}.
\label{inv_LL}
\end{eqnarray}
$J^{(d_i)}_{LL}$ and $J^{(u_i)}_{LL}$ contribute to the $i$-th down-
and up-type quark (C)EDMs. Since chirality of quark is flipped in the
quark (C)EDM operators (see Eq.~(\ref{eff_cp})), the invariants are
proportional to the corresponding Yukawa coupling constants.  The
relative phases between the CKM matrix and $\delta_{LL}^q$ contribute
to the invariants.  In the MSUGRA model, non-zero values for
$(\delta_{LL}^{q})_{ij}$ are induced by the RG effect, and they turn
out to be proportional to $(Y_u)_{ij}$ or $(Y_d)_{ij}$.  However,
those flavor structures are not able to contribute to the Jarlskog
invariants.  This implies that only non-minimal flavor structures
generate the quark (C)EDMs via the new invariants, as stressed above.

\begin{figure}[t]
\begin{tabular}{cc}
\includegraphics[scale=1.0]{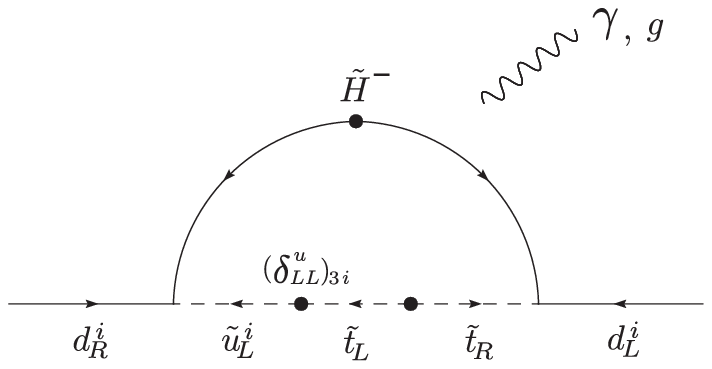} &
\includegraphics[scale=1.0]{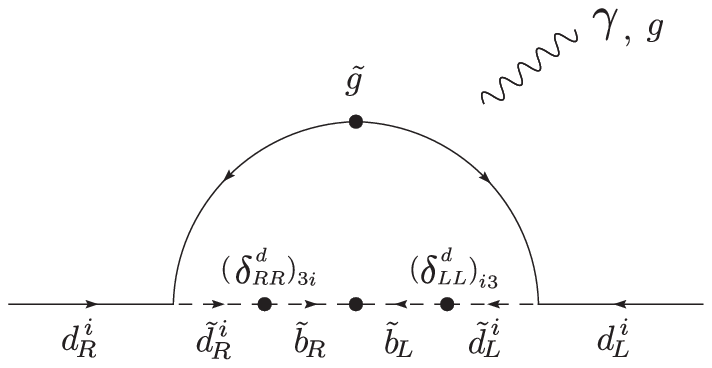}
\end{tabular}
\caption{\label{fig1} Left: Higgsino contribution to the $i$-th
  down-quark (C)EDM in a case of non-vanishing left-handed squark
  mixing, $(\delta_{LL}^q)_{3j}\ne 0$. Right: Gluino contribution to
  the $i$-th down-quark (C)EDM in a case of non-zero values for both
  left- and right-handed squark mixings, $(\delta_{LL}^q)_{i3}\ne 0$
  and $(\delta_{RR}^d)_{3i}\ne 0$.}
\end{figure}

Higgsino diagrams may contribute to the (C)EDMs at one-loop level in
the case of $(\delta_{LL}^q)_{ij}\ne 0$ (see Fig.~\ref{fig1}), since
$\delta_{LL}^q$ is sandwiched between the Yukawa coupling constants in the
invariants \cite{Endo:2003te}.  The down-quark CEDM and EDM are larger
than those of up quark for moderate $\tan\beta$, since $J^{(u)}_{LL}$
is suppressed by the square of down-type quark Yukawa coupling
constants.  In the limit of degenerate SUSY particle masses, it is found
that the down- and strange-quark EDMs and CEDMs are given as 
\beq
\left\{\frac{\left(d_{d_i}\right)_{\tilde{H}^{\pm}}}{e}
,~(d_{d_i}^c)_{\tilde{H}^{\pm}}\right\}
=
-\frac{\alpha_2}{4\pi}\left(\frac{m_{d_i}}{m_W^2}\right)
\left(\frac{m^{2}_{t}}{m_{\tilde{q}}^2}\right)
\left(\frac{\mu A_{t}}{{m}_{\tilde{q}}^2}\right)\tan\beta
\bigg[\im\left\{(\delta^{u}_{LL})_{i3}V_{3i}\right\}\bigg]
\left\{\frac{1}{30},~\frac{1}{40}\right\},
\label{higgsinoedm}
\eeq
where $i=1,2$.  The (C)EDMs are suppressed by the external light quark
masses, as expected from the invariants (\ref{inv_LL}).  For
$|(\delta_{LL}^q)_{13}|\sim \lambda^3$, $m_{SUSY}=500\,\rm{GeV}$ and
$\tan\beta=10$, $d_d/e$ and $d_d^c$ are of order $\sim
10^{-28}~{\rm{cm}}$.

When $(\delta_{RR}^d)_{ij}\ne 0$ or $(\delta_{RR}^u)_{ij}\ne 0$, the new
Jarlskog invariants are
\begin{eqnarray}
J^{(d_i)}_{RR}= {\rm Im} \left\{Y_uf_d\delta_{RR}^d \right\}_{ii},
\nonumber\\
J^{(u_i)}_{RR}= {\rm Im} \left\{Y_df_u\delta_{RR}^u\right\}_{ii}.
\label{inv_RR}
\end{eqnarray}
In the above cases, no diagram contributes to the quark (C)EDMs at
one-loop level, since the mass insertions appear in the right-hand
side of the Yukawa coupling constants in the invariants.  The first
non-vanishing contribution arises from higher-order effects (two-loop
effects), and the quark (C)EDMs are suppressed by loop factors.
However, the (C)EDMs may be enhanced by the Yukawa coupling constant
of the heaviest generation via the CKM and right-handed squark
mixings, that partially compensates the loop suppression.  In the
next section, we will show that the non-holomorphic Yukawa coupling
for $\bar{u}_{Li} d_{Rj} H^+$ vertex is generated at one-loop level
and is proportional to $f_{d_i}(\delta_{RR}^d)_{ij}$ in the cases of
$(\delta_{RR}^d)_{ij}\ne 0$. Thus, the charged-Higgs loop contributes
to the down-quark (C)EDM at two-loop level, providing a contribution
proportional to $J^{(d)}_{RR}$.

In a case of  both left- and right-handed squarks have mixings,
the (C)EDMs are derived at one-loop level and are also enhanced
by the heaviest quark masses. The new Jarlskog invariants are
\begin{eqnarray}
J^{(d_i)}_{LR}= {\rm Im} \left\{\delta_{LL}^qf_d\delta_{RR}^d
\right\}_{ii},
\nonumber\\
J^{(u_i)}_{LR}= {\rm Im} \left\{\delta_{LL}^qf_u\delta_{RR}^u\right\}_{ii}.
\label{inv_LR}
\end{eqnarray}
The above invariants are proportional to only one Yukawa coupling
constant, and the relative phases of the left- and right-handed squark
mixings contribute to them. 

The gluino diagrams in Fig.~\ref{fig1} contribute to the quark
(C)EDMs, being proportional to the above $J^{(d_i,u_i)}_{LR}$ invariants
\footnote{The bino diagrams similar to Fig.~\ref{fig1} also contribute
  to the quark (C)EDMs, being proportional to the above
  $J^{(d_i,u_i)}_{LR}$ invariants. However, they are always sub-dominant 
  compared to the gluino ones.
}.
The down- and strange-quark EDMs and CEDMs are given as
\begin{eqnarray}
\left\{\frac{\left(d_{d_i}\right)_{\tilde{g}}}{e}
,~(d_{d_i}^c)_{\tilde{g}}\right\} &=&
- \frac{\alpha_s}{4\pi}\frac{m_{b}\mu\tan\beta}{m^3_{\tilde{q}}}
\bigg[{\rm {Im}}\left\{(\delta_{LL}^{d})_{i3}\,(\delta_{RR}^{d})_{3i}\right\}\bigg]
\left\{\frac{4}{135},~\frac{11}{180}\right\},
\label{dsap}
\end{eqnarray}
where $i=1,2$.  In the above expression we have set the gluino mass
$m_{\tilde{g}}$ to be equal to the squark masses $m_{\tilde{q}}$.  The
down-quark EDM reaches to the $10^{-(25-26)}~e\, {\rm cm}$ level even for
$m_{SUSY}=500$GeV, $(\delta_{LL}^{d})_{13}\sim
(\delta_{RR}^{d})_{31}\sim \lambda^3$ and $\tan\beta=10$.  Thus, in
this case, the current experimental bounds on the hadronic EDMs give
constraints on the model parameters.  SUSY GUTs predict both
left-handed and right-handed squark mixings, and the related one-loop
induced EDMs have been studied in many papers \cite{gutedm}.

In the next section, we will introduce the non-holomorphic Yukawa
couplings of the charged-Higgs boson and evaluate the down-quark
(C)EDM at two-loop level when $(\delta_{LL}^q)_{ij}\ne0$ or
$(\delta_{RR}^d)_{ij}\ne0$.  Other cases where
$(\delta_{RR}^u)_{ij}\ne0$ are remanded to our future works.  The
up-quark (C)EDM from the charged-Higgs loop is proportional to
$\tan\beta$ while $J^{(u)}_{RR}$ is proportional to $\tan^2\beta$.
This implies that we need a more complete analysis since other
two-loop diagrams may dominate over it.

\section{Non-holomorphic Yukawa Interactions and Quark (C)EDMs}

\begin{figure}[t]
\begin{tabular}{cc}
\includegraphics[scale=1.0]{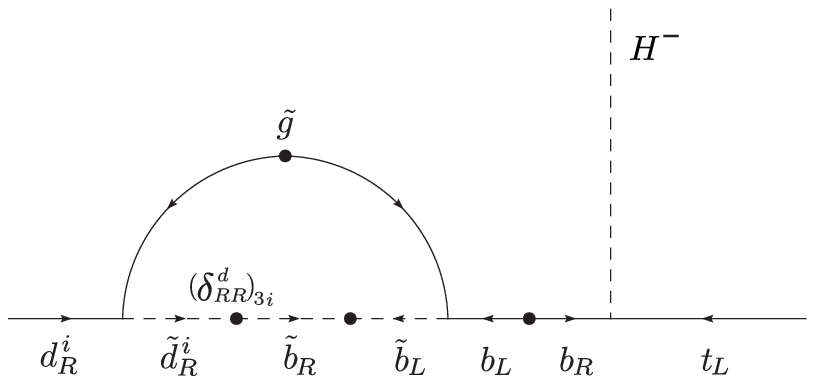} &
\includegraphics[scale=1.0]{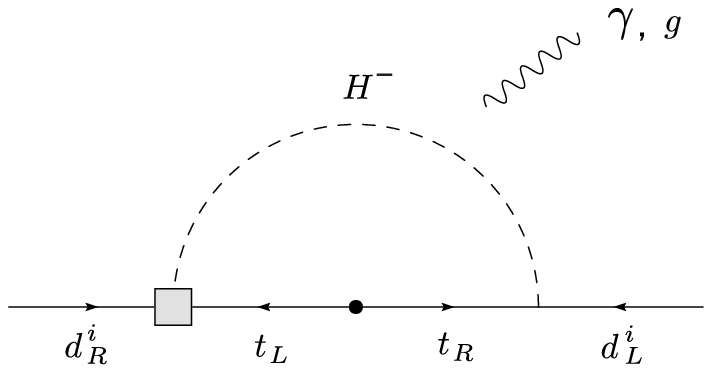}
\end{tabular}
\caption{\label{fig2} Left: Leading contribution to the effective
  $\bar{t}_L d^i_R H^+$ coupling in Eq.~(\ref{MIA:EW:GamR1}) in a
case of $(\delta_{RR}^d)_{i3}\ne0$.
  Right:~Charged-Higgs mediated diagrams contributing to the
  (C)EDMs  in a
case of $(\delta_{RR}^d)_{i3}\ne0$. }
\end{figure}

In the present paper, we are mainly interested in possible
CP-violating effects arising from the Higgs sector through
non-holomorphic Yukawa interactions that we are going to discuss.
When $(\delta_{RR}^d)_{3i}\ne0$ $(i=1,2)$, the effective $\bar{t}_L
d^i_R H^+$ coupling is derived from the wave function renormalization
of right-handed down-type quarks as shown in the Feynman diagram of
Fig.~\ref{fig2}, and it is enhanced by the bottom-quark mass. The
charged-Higgs loop contributes to the down- and strange-quark (C)EDMs
that result to be proportional to the Jarlskog invariants
$J^{(d,s)}_{RR}$, as expected.  Given that the above contributions are
generated by non-holomorphic Yukawa interactions, they do not decouple
even when the SUSY particle masses are very heavy \footnote{
  As recently shown in Ref.~\cite{kl2,kpnn}, charged-Higgs boson can induce
  other very interesting non-standard phenomena in the $K$ meson
  system: violations of lepton universality which can reach the $1\%$
  level in the $\Gamma(K\rightarrow e\nu)/\Gamma(K\rightarrow \mu\nu)$
  ratio \cite{kl2} and large deviation from the SM expectation in the
  branching ratios of $K\rightarrow\pi\nu\bar\nu$ \cite{kpnn}.
  .
}.

Following the procedure of Refs.~\cite{IR2,Foster}, in order to compute
all the non-decoupling effects at large $\tan\beta$ one needs to: {\it
  i}) evaluate the effective dimension-four operators appearing at 
one-loop level which modify the tree-level Yukawa Lagrangian; {\it
  ii}) expand the off-diagonal mass terms in the squark sector by
means of the mass-insertion approximation; {\it iii}) diagonalize the
quark mass terms and derive the effective interactions between quarks
and heavy Higgs fields.

The effective interaction of $\bar{t}_L d^i_R H^+$ and $\bar{t}_R d^i_L H^+$ 
is parametrized as 
\bea
\cL_{H^\pm}^{\rm eff}&=& 
\frac{g }{\sqrt{2}m_W}\frac{m_{t}}{\tan\beta}V_{3i}C_{L}^{H^+}
~\bar{t}_R d^i_L H^+
\no\\
&+&\frac{g }{\sqrt{2}m_W} m_{d_i} \tan\beta
V_{3i}C^{H^+}_{R}
~\bar{t}_L d^i_R H^+
~+~ {\rm h.c.}.
\label{eq:effH}
\eea The dominant corrections to the charged-Higgs vertex arise from
gluino exchange, and the effective charged-Higgs coupling constants
are given by
\beq
C_L^{H^+}
=
(1-\epsilon_s \tanb)+
\epsilon_{f}\frac{\left(\dll\right)_{3i}}{V_{3i}}\tanb,
\label{MIA:EW:Del}
\eeq
\beq
C_R^{H^+}
=
 \frac{1}{\left(1+\epsilon_s\tan\beta\right)}
\left(1
-\frac{\epsilon_{f}\tanb}{\left(1+\epsilon_s\tan\beta\right)}
\frac{\left(\dll\right)_{3i}}{V_{3i}}
+\frac{\epsilon_{f}\tanb}{\left(1+\epsilon_s\tan\beta\right)}
\frac{m_b}{m_{d_i}}\frac{\left(\drr\right)_{3i}}{V_{3i}}
\right).
\label{MIA:EW:GamR1}
\eeq
The sub-dominant Higgsino contributions are neglected.  They become
relevant only for very large values of $\tan\beta$, {\it i.e.}
$\tan\beta\gsim 40$. Here, $V$ refers to the renormalized CKM matrix at
one-loop level. When $\tan\beta$ is quite large, the radiative
corrections to the CKM matrix are not negligible. The elements of the
renormalized CKM matrix, which are relevant in the following
discussion, are given from the tree-level ones ($V^{\rm tree}$) as
\begin{eqnarray}
V_{3i}=V^{\rm tree}_{3i} +\frac{\epsilon_{f} \tan\beta}{1+\epsilon_s\tan\beta}
(\delta_{LL}^d)_{3i}.
\end{eqnarray}
In the above equations, the $\epsg$ and $\epsilon_{f}$ terms are
given by
\begin{eqnarray}
\epsg=\frac{\alpha_s}{3\pi}\frac{\mu}{m_{\tilde{q}}}, &&
\epsilon_{f}=\frac{\alpha_s}{9\pi}\frac{\mu}{m_{\tilde{q}}}.
\label{MIA:mdb:aux2}
\end{eqnarray}
Here, we assumed that the squark masses $m_{\tilde{q}}$ are equal to
the gluino mass $m_{\tilde{g}}$.

It is noticed that the effective coupling in Eq.~(\ref{MIA:EW:GamR1})
contains terms not suppressed by the $m_{d_i}(i=1,2)$ factor. Thanks
to the right-handed squark mixing terms, we can always pick up the bottom
mass $m_b$ instead of $m_{d}$ or $m_{s}$.  We stress that, in the MFV
framework, it is not possible to get such a Yukawa enhancement.

The (C)EDMs of down and strange quarks are induced by the one-loop
exchange of the charged-Higgs boson and top quark, as it is shown in
Fig.~\ref{fig2}. By making use of the effective Lagrangian in
Eq.~(\ref{eq:effH}), they are evaluated as
\bea
\left\{\frac{\left(d_{d_i}\right)_{H^{\pm}}}{e},~
\left(d_{d_i}^c\right)_{H^{\pm}}\right\}\!\!
&=&
\frac{\alpha_2}{4\pi}|V_{3i}|^{2}
\left(\frac{m_{d_i}}{m_W^2}\right)\left(\frac{m^{2}_{t}}{m^{2}_{H^{\pm}}}\right)
\,\,\bigg[\im\left\{C_{L}^{H^+}C_{R}^{H^+*}\right\}\bigg]
\nonumber\\
&&\times
\left\{F_{7},~F_8\right\}(y_{tH})\,,
\eea
where $y_{tH}=m^{2}_t/m^{2}_{H^{\pm}}$ and the loop functions
$F_{7}(x)$ and $F_{8}(x)$ are given by
\be
F_{7}(x) = \frac{3-5x}{12(1-x)^{2}}-\frac{3x-2}{6(1-x)^{3}}\log(x),~\qquad
F_{8}(x) = \frac{3-x}{4(1-x)^{2}}+\frac{\log(x)}{2(1-x)^{3}}~.
\label{floop}
\ee
In the case of $(\delta_{RR}^d)_{i3}\ne0$ $(i=1,2)$, the charged-Higgs
contributions to the (C)EDMs are given as 
\bea \left\{\frac{\left(d_{d_i}\right)_{H^{\pm}}}{e},~
  \left(d_{d_i}^c\right)_{H^{\pm}}\right\}\!\!  
&=&
\!\!-\frac{\alpha_2}{4\pi}
\bigg[\im\left\{V_{3i}^\star(\delta^{d}_{RR})_{3i}\right\}\bigg]
\left(\frac{m_{b}}{m_W^2}\right)\!\left(\frac{m^{2}_{t}}{m^{2}_{H^{\pm}}}\right)
\!\frac{\epsilon_{f}(1\!-\!\epsilon_s\tan\beta)\tan\beta}{(1\!+\!\epsilon_s\tan\beta)^2}
\nonumber\\
&&\times \left\{F_{7},~F_8\right\}(y_{tH})~,
\label{main}
\eea
while in the case of $(\delta_{LL}^d)_{i3}\ne0$ $(i=1,2)$
\bea
\left\{\frac{\left(d_{d_i}\right)_{H^{\pm}}}{e},~
\left(d_{d_i}^c\right)_{H^{\pm}}\right\}\!\!
&=&
-\frac{\alpha_2}{4\pi}\bigg[\im\left\{(\delta^{d}_{LL})_{i3}V_{3i}\right\}\bigg]
\left(\frac{m_{d_i}}{m_W^2}\right)\!\left(\frac{m^{2}_{t}}{m^{2}_{H^{\pm}}}\right)
\!\frac{2 \epsilon_{f}\tan\beta}{(1\!+\!\epsilon_s\tan\beta)^2}
\nonumber\\
&&
\times\left\{F_{7},~F_8\right\}(y_{tH}).
\label{subd}
\eea
The above contributions are proportional to $J^{(d_i)}_{RR}$ and
$J^{(d_i)}_{LL}$, respectively.  Making a comparison with the
contributions in Eq.~(\ref{main}), it is observed that the effects in
Eq.~(\ref{subd}) are suppressed by a $m_{d_i}/m_b$ factor, as
expected.

In principle, the exchange of the neutral-Higgs bosons, $H^0$ and
$A^0$, could also contribute to the (C)EDMs. Their contributions are
enhanced by a $\tan^{\!4}\beta$ factor, due to the nature of the
non-holomorphic Yukawa couplings \cite{Babu:1999hn}, while they suffer
from a suppression of order $(m_b/m_t)^{2}$, compared to the
charged-Higgs case. However, these effects
arise at the three-loop level and are additionally reduced by a
cancellation between the $A^0$ and $H^0$ contributions (due to the
mass degeneration of $A^0$ and $H^0$ at tree level).  We found that 
the neutral-Higgs effects to the (C)EDMs are largely sub-dominant
compared to the charged-Higgs ones even in the most favorable
circumstance, namely for large $\tan\beta$ values and for moderate
mass splitting between $H^0$ and $A^0$ ($\delta m/m_{A^0}\sim 10\%$).

\begin{figure}[t]
\begin{tabular}{cc}
\includegraphics[scale=0.4]{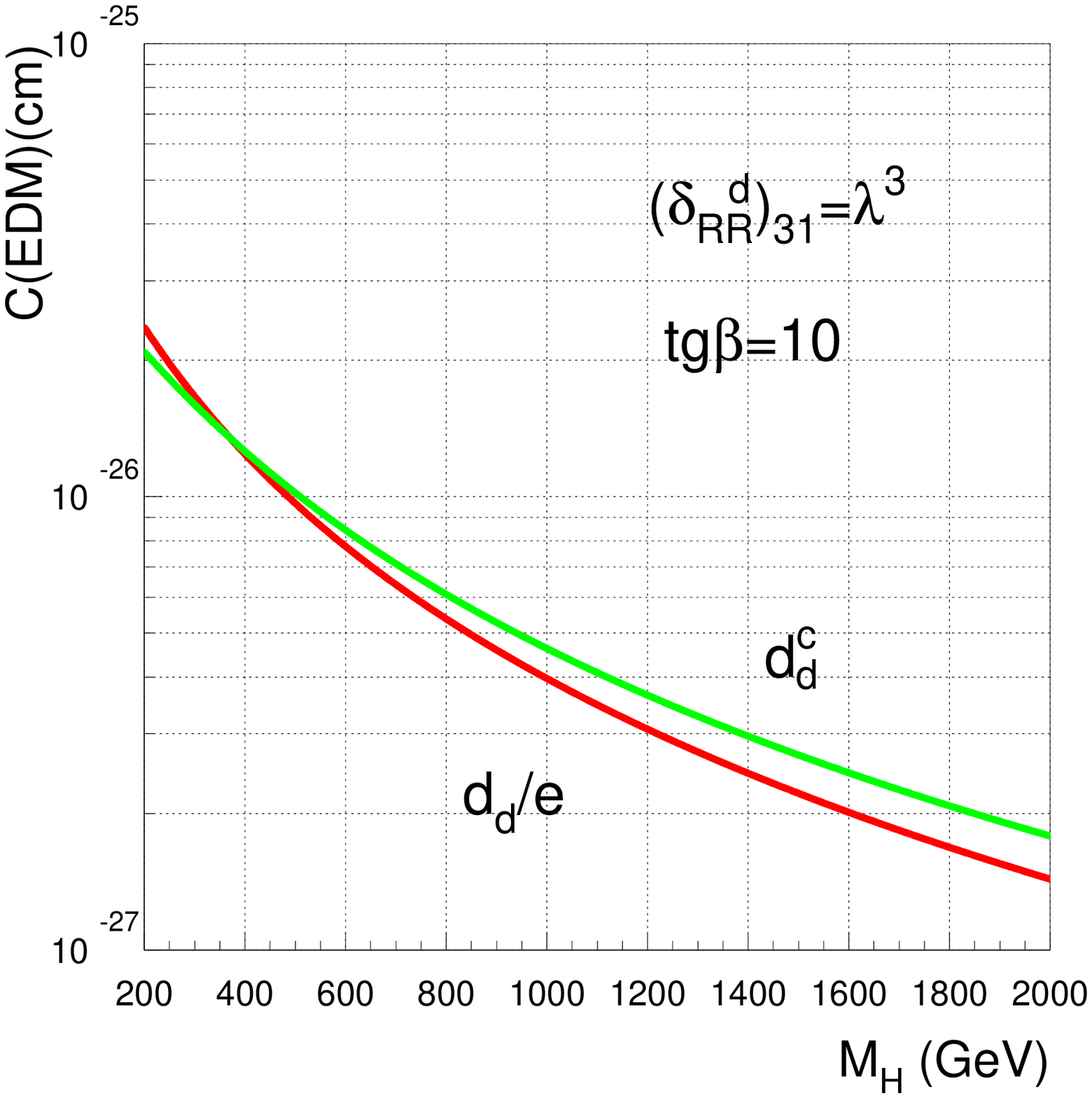} &
\includegraphics[scale=0.4]{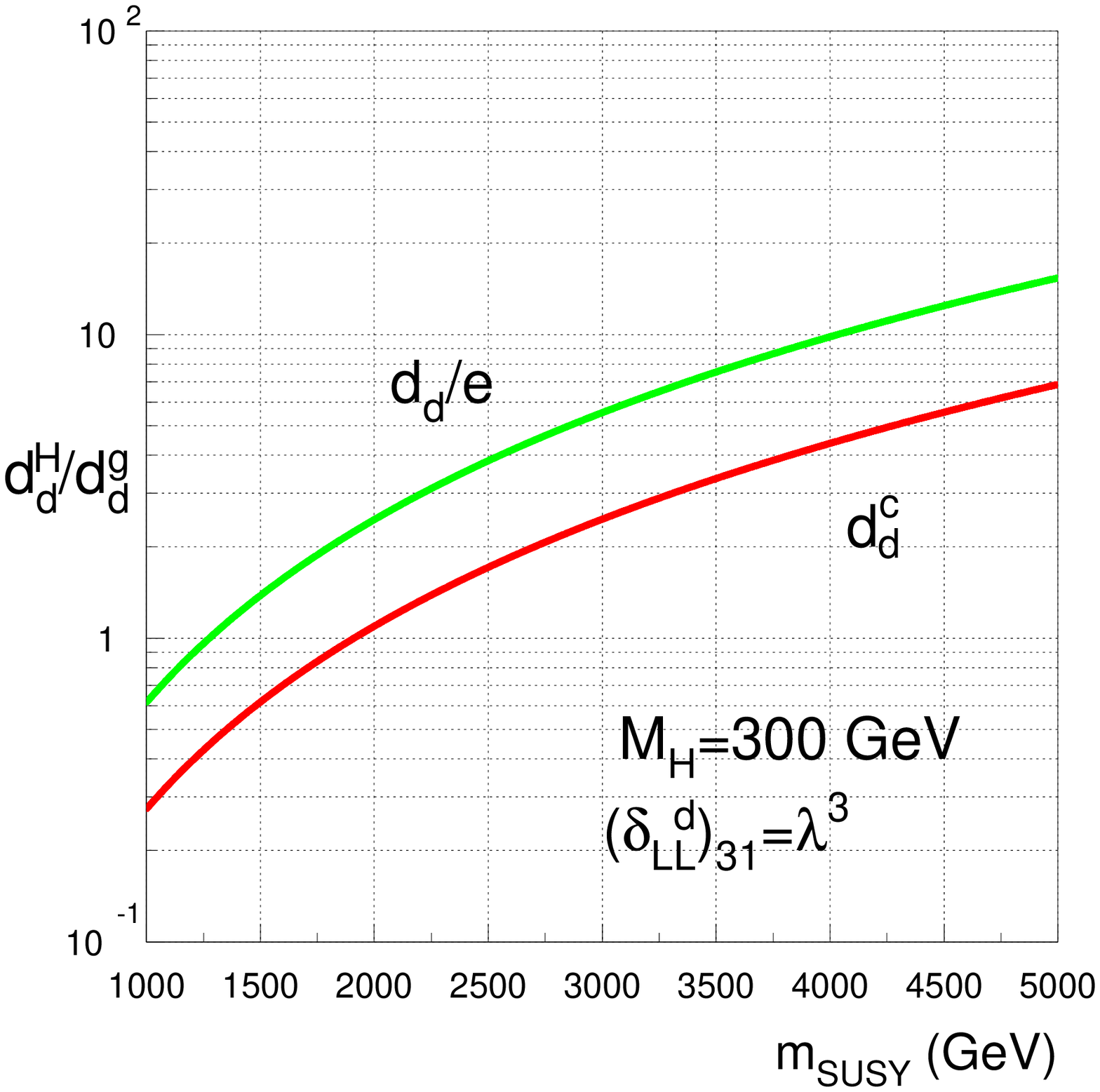}
\end{tabular}
\caption{\label{fig3}Left: Down-quark EDM and CEDM as functions of the
  charged-Higgs mass $m_{H^{\pm}}$ for $\tan\beta=10$. The (C)EDM
  scales as $\tan\beta/10$. 
   In this figure, a CKM-type mixing is assumed for $\delta^{d}_{RR}$,
  {\it i.e.} $|(\delta^{d}_{RR})_{31}|=\lambda^3$.
   Right:~Ratio of charged-Higgs and gluino
  induced down-quark (C)EDM as a function of a common SUSY particle
  mass  $m_{SUSY}$, setting the Higgs mass to $m_{H^{\pm}}=300
  \rm{GeV}$.  We assume ${(\delta^{d}_{LL})_{13}}=-{V_{31}^\star}$.
}
\end{figure}

\subsection*{Numerical analysis}

The sensitivity of the down-quark (C)EDM to the charged-Higgs mediated
effects is illustrated on the left of Fig.~\ref{fig3}.  In this
figure, we assume $|(\delta^{d}_{RR})_{31}|=\lambda^3$. From this figure,
$(d^{c}_{d})_{H^{\pm}}$ and $(d_{d}/e)_{H^{\pm}}$ can be of order
$10^{-(26-27)}~{\rm cm}$ for $\tan\beta=10$ and $m_{H^{\pm}}\in (0.2,
2)\,\rm{TeV}$\footnote{
  In the following, we take $sign(\mu A)<0$, as it is
  preferred by the $(g-2)_{\mu}$ and $b\rightarrow s\gamma$
  experimental results.
}.
On the other hand, the current experimental upper bounds on the EDMs
of neutron, $|d_{n}|<3.0 \times 10^{-26}~e\,{\rm cm}$
\cite{Baker:2006ts}, and of $^{199}$Hg atom, $|d_{\rm Hg}| < 2.1\times
10^{-28}~e\, {\rm cm}$ \cite{Romalis:2000mg}, constrain the down-quark
EDM and CEDM as $|d^{c}_{d}|$ and $|d_{d}|/e\lsim 10^{-26}~{\rm cm}$,
barring accidental cancellations among contributions
\cite{Pospelov:2005pr}\cite{Hisano:2004tf}. The charged-Higgs mediated
effects to the (C)EDM are very close to the actual experimental
sensitivities and will be strongly probed with the expected future
experimental resolutions.

The strange-quark CEDM is easily obtained by means of the down-quark
CEDM in the following way,
\beq 
\frac{d^c_s}{d^c_d}= \frac{\im\left\{V^\star_{32}(\delta^{d}_{RR})_{32}\right\}}
{\im\left\{V^\star_{31}(\delta^{d}_{RR})_{31}\right\}}.
\label{dsd}
\eeq
The neutron EDM might receive a sizable contribution from the
strange-quark CEDM as $d_n\sim 0.1 ed^c_s$ \cite{scedm,Hisano:2004tf},
while the QCD uncertainties are still not well enough under
control \cite{Pospelov:2005pr}.  When the neutron EDM bound gives a
significant constraint on the strange-quark CEDM, it is transmitted as
a bound on $(\delta^{d}_{RR})_{32}$.

When the SUSY scenario contains only the left-handed squark mixings,
the expected values for the down-quark EDM and CEDM are reduced by a
$m_d/m_b$ factor compared to the case of Fig.~\ref{fig3}.  They can
reach to the level of $10^{-28}~(e){\rm cm}$, which is comparable to
the one-loop Higgsino contribution, given in Eq.~(\ref{higgsinoedm}).

A non-trivial constraint on the charged-Higgs mass is extracted from
$\Gamma(B \to X_s \gamma)$.  It is well known that, the charged-Higgs
and chargino contributions to this observable must have opposite sign
in order to fulfill its precise experimental determination, when the
chargino and charged Higgs masses are not heavy enough to be
decoupled.  The relative sign between these two amplitudes is
proportional to $ {\rm sign}(\mu A_t)$, where $A_t$ is the trilinear
soft-breaking term appearing in the stop mass matrix.  However, if the
SUSY particles are heavy (above the TeV scale), the chargino
contributions are not so effective, and $\Gamma(B \to X_s \gamma)$
gets the dominant effect by the charged-Higgs contributions. In such
cases, it turns out that $m_{H^{\pm}}\gsim 300 \rm{GeV}$ \cite{bsg}.
This lower bound on the charged-Higgs mass may be more relaxed (below
$m_{H^{\pm}}\sim 300 \rm{GeV}$) in the large $\tan\beta$ regime, since
sizable higher-order (non-holomorphic) effects interfere destructively
with the leading contributions \cite{bsg2}.

An interesting feature of the charged-Higgs contributions to the
(C)EDMs is the slow decoupling for $m_{H^{\pm}} \to \infty$, in
contrast to the one-loop gluino or Higgsino contributions
(Fig.~\ref{fig1}).  Being generated by effective one-loop diagrams,
the $F_{7,8}(y_{tH})$ loop functions develop a large logarithm so that
the (C)EDMs behave as $(d^{(c)}_{q})_{H^{\pm}}\sim y_{tH}
\log(y_{tH})$ for $y_{tH}\!=\!m_t^2/m_{H^{\pm}}^2\!\ll 1$.  This
property is clearly shown in Fig.~\ref{fig3}, where to one order of
magnitude increase of the Higgs mass corresponds only a 10 factor
reduction of the (C)EDMs.  As a result, even for moderate $\tan\beta$
values and charged-Higgs masses $m_{H^{\pm}}< 2 ~\rm{TeV}$, the
down-quark (C)EDM remains at a potential visible level, {\it i.e.}
$(d_{d}/e)_{H^{\pm}}$ and $(d^{c}_{d})_{H^{\pm}} \sim\mathcal O
(10^{-(26-27)})~{\rm cm}$ for $|(\delta^{d}_{RR})_{31}|=\lambda^3$.

Next, let us compare the two-loop charged-Higgs contributions
in Eq.~(\ref{main}) to the one-loop gluino ones
(Fig.~\ref{fig1}), in the case of $(\delta^{d}_{RR})_{3i}\ne 0$.
Assuming the simple case of degenerate SUSY particle masses, one gets
\begin{eqnarray}
\left\{\frac{\left(d_{d_i}\right)_{H^{\pm}}}{\left(d_{d_i}\right)_{\tilde{g}}},~
\frac{\left(d_{d_i}^c\right)_{H^{\pm}}}{\left(d_{d_i}^c\right)_{\tilde{g}}}\right\}
\!\!
&\simeq&
\frac{\alpha_2}{9\pi}
\left(\frac{m^{2}_{t}}{m^{2}_{W}}\right)
\left(\frac{m^{2}_{\tilde{q}}}{m^{2}_{H^{\pm}}}\right)
\frac{
{\im}\left\{(\delta_{LL}^{d})_{i3}\,(\delta_{RR}^{d})_{3i}\right\}
}
{\im\left\{V_{3i}^{\star}(\delta^{d}_{RR})_{3i}\right\}}
\nonumber\\
&&\times\left\{\frac{135}{4}F_{7},~\frac{180}{11}F_8\right\}(y_{tH}).
\label{hglu}
\end{eqnarray}
with $F_7\in (-0.24, -1.41)$ and $F_8\in (-0.21, -1.76)$ when
$m_{H^{\pm}}\in(0.2,2)\rm{TeV}$. Here, we assumed moderate $\tan\beta$
values and ignored higher-order terms of $\alpha_s$. The plot on the
right of Fig.~\ref{fig3} shows the ratio between charged-Higgs and
gluino contributions to the down-quark (C)EDM as a function of the
squark mass.  In the figure, we take
${(\delta^{d}_{LL})_{13}}=-{V_{31}^\star}$.  The charged-Higgs
mediated effects do not vanish even in the limit of heavy squarks and
gluinos, provided the Higgs sector remains relatively light. As a
result, the two-loop Higgs mediated (C)EDMs are larger than the
one-loop gluino mediated ones if $m_{SUSY}\gsim (1-2)\rm{TeV}$ when
$m_{H^{\pm}}\sim 300 \rm{GeV}$.

Noteworthy, it is found from Eq.~(\ref{hglu}) that the ratios of
charged-Higgs and gluino contributions are independent of the $\mu$
sign and also on $\tan\beta$, for moderate $\tan\beta$ values.  When
the non-zero values of ${(\delta^{d}_{LL})_{i3}}$ come from the
RG effect due to the top-quark Yukawa coupling as in the
MSUGRA model, it is proportional to $-{V_{3i}^{*}}$. Thus, the one-
and two-loop effects are constructive.

When the SUSY framework contains the only flavor structures in the
left-handed squark mass terms, the ratio of the charged-Higgs and
Higgsino contributions to the down- and strange-quark (C)EDMs are
\bea
\left\{\frac{\left(d_{d_i}\right)_{H^{\pm}}}{\left(d_{d_i}\right)_{\tilde{H}^{\pm}}},~
  \frac{\left(d_{d_i}^c\right)_{H^{\pm}}}{\left(d_{d_i}^c\right)_{\tilde{H}^{\pm}}}\right\}\!\!  
&\simeq& 
\frac{2\alpha_s}{9\pi}
\left(\frac{m^{2}_{\tilde{q}}}{m^{2}_{H^{\pm}}}\right)
\left(\frac{A_t}{m_{\tilde{q}}}\right)^{-1}
\frac{\im\left\{(\delta^{d}_{LL})_{i3}V_{3i} \right\}}{\im\left\{(\delta^{u}_{LL})_{i3}V_{3i} \right\}}
\nonumber\\
&& \times \left\{30F_{7},~40F_8\right\}(y_{tH})\,.
\label{hchar}
\eea
Reminding that $\delta_{LL}^u=V\delta_{LL}^dV^\dagger$
because of the SU(2)$_L$ symmetry, it turns out that
$(\delta^{u}_{LL})_{13}\sim (\delta^{d}_{LL})_{13}+\lambda(\delta^{d}_{LL})_{23}$
and $(\delta^{u}_{LL})_{23}\sim (\delta^{d}_{LL})_{23}-\lambda(\delta^{d}_{LL})_{13}$.
From Eq.~(\ref{hchar}) it is straightforward to check that,
barring accidental cancellations among contributions arising from 
different flavor structures, the charged-Higgs effects to down- and 
strange-quark (C)EDMs dominate over the chargino ones
even if $m_{SUSY}$ is few times larger than  $m_{H^{\pm}}$.
Thus, the charged-Higgs contribution should be included in the case that 
only the left-handed squark matrix has non-minimal flavor structures.

\section{Conclusions and Discussion}

In this paper, we have discussed the quark EDMs and CEDMs in the MSSM
with GFV terms in the squark mass matrices.  In the SM, where the only
source of CP violation comes from the CKM matrix, the EDMs are
strongly suppressed by high powers of the Yukawa coupling constants.  This
implies that the EDMs may be very sensitive to non-minimal flavor  
structures of theories beyond the SM, such as the MSSM.  In
the MSSM, the first contributions to the (C)EDMs are generated at 
one-loop level by exchange of Higgsino/squarks or gluino/squarks.
While in the former case, non-vanishing left-handed squark mixings are
enough to generate the EDMs, in the latter case both left and
right-handed squarks mixings are required.

In this work, we have pointed out that new two-loop contributions
mediated by the charged-Higgs exchange may generate the quark (C)EDMs.
In particular, we have evaluated the down- and strange-quark EDMs and
CEDMs.  Interesting enough, the above contributions do not decouple
even for heavy squarks and gluino.  They are proportional to
$\tan\beta$ for moderate $\tan\beta$, similar to the one-loop
contributions. These new contributions may dominate over the one-loop
contributions in the case of non-vanishing right-handed down-type squark
mixing, even when the squark and gluino masses are around the TeV
scale. Furthermore, the down-quark EDM and CEDM may be very close to
the actual experimental sensitivities, since $d_{d}/e$ and $d^{c}_{d}
\sim\mathcal O (10^{-(26-27)})~{\rm cm}$ for
$|(\delta^{d}_{RR})_{31}|=\lambda^3$. They will be strongly probed
with the expected future experimental resolutions.

In the case that the only left-handed squark mixings are
non-vanishing, the two-loop contributions may dominate over the
one-loop Higgsino ones even when SUSY particle masses are lighter than
1~TeV.  Thus, the two-loop contribution cannot be neglected. The
down-quark EDM and CEDM may reach the level of $\sim\mathcal O
(10^{-28})~(e){\rm cm}$ for $|(\delta_{LL}^q)_{13}|\sim\lambda^3$, and
they might be also covered in the future experiments.

At large $\tan\beta$, the observable EDMs of neutron and heavy atoms
receive contributions not only from the (C)EDMs of the constituent
particles, such as electrons and quarks, but also from CP-odd
four-fermion operators through the exchange of neutral-Higgs bosons
\cite{Demir:2003js}.  These effects would be sensitive to CP phases in
the flavor-diagonal $F$-term SUSY breaking terms.  We also noticed
that within the GFV framework, the phases in the squark mass matrices
might also provide a contribution to the CP-odd four-fermion operators.
This happens when both left-handed and right-handed squarks have
mixing.  However, we found that these contributions are smaller than
those arising from the charged-Higgs exchange, at least for moderate
$\tan\beta$.

Finally, we discuss a correlation between the hadronic EDMs and FCNC
processes. In Ref.\cite{bphiks}, it has been pointed out that, within
some SUSY models, ${d}^c_s$ and the deviation from the SM prediction
of the CP asymmetry in $B\rightarrow \phi K_S$ ($\delta S_{\phi K_s}$)
are correlated. Non-vanishing right-handed down-type squark mixing
${(\delta^{d}_{RR})_{23}}$ may induce sizable $\delta S_{\phi K_s}$
via the gluino penguin diagram.  The effective Lagrangian relevant for
the penguin diagrams is ${\cal L^{\rm eff}}= {g_s m_b}/{(16\pi^2)}
C^{L}_{8} \bar{s}_L(G\cdot \sigma) b_R +{g_s m_b}/{(16\pi^2)}
C^{R}_{8} \bar{s}_R(G\cdot \sigma) b_L$, and the right-handed squark
mixing contributes to $C^{R}_{8}$. If we additionally require that
${(\delta^{d}_{LL})_{23}}\ne0$, such as in the MSUGRA model, it turns
out that ${d}^c_s$ and $\delta S_{\phi K_s}$ are correlated, since
\begin{eqnarray}
\left({d}^c_s\right)_{\tilde{g}} &=& \frac{m_b}{8\pi^2} 
\frac{11}{21}
{\rm {Im}}\left\{( \delta_{LL}^{d})_{32} \left(C^{R}_{8}\right)_{\tilde{g}} \right\}\,,
\label{massin}
\end{eqnarray}
up to the QCD correction.

A similar correlation also appears when the the charged-Higgs
mediation effect is the dominant source for $\delta S_{\phi K_s}$. 
In the case of $(\delta_{RR}^{d})_{23}\ne0$, 
\begin{eqnarray}
\left({d}^c_s\right)_{H^{\pm}} &=& \frac{m_b}{8\pi^2}
{\mathrm{Im}}\left\{V_{32} \left(C^{R}_{8}\right)_{H^{\pm}}\right\}.
\end{eqnarray}
Then, in the charged-Higgs mediated case, $\delta S_{\phi K_s}$ is
also constrained by the hadronic EDMs, when the hadronic EDMs receive
sizable contributions from the strange-quark CEDM.  A similar
discussion applies also to $b\rightarrow s/d+\gamma$ and so on. A
detailed analysis of these last effects is remanded to our future
works. 
\vskip 0.5in 
\vbox{
\noindent{ {\bf Acknowledgments} } \\
\noindent  
The work of J.H. is supported in part by the Grant-in-Aid for Science
Research, Ministry of Education, Science and Culture, Japan
(No.~1554055 and No.~18034002).  Also, that of M.N.
is supported in part by JSPS.

}

\end{document}